\documentstyle[12pt]{article} 
\begin{document}

\begin{flushright}
hep-ph/9710418 \\
DPNU-97-49 \\
October 1997 
\end{flushright}

\vspace{10mm}

\begin{center}
{\large  \bf 
Large Lepton Flavor Mixing \\
and \\
${\bf E_6}$-type Unification Models }
\end{center}

\vskip 10mm

\begin{center} 
Naoyuki HABA$^1$ and Takeo MATSUOKA$^2$  \\

\vspace{5mm}

{\it 
${}^1$Faculty of Engineering, Mie University \\
           Mie, JAPAN 514 \\
${}^2$Department of Physics, Nagoya University \\
           Nagoya, JAPAN 464-01 
}
\end{center}

\vspace{30mm}

\begin{abstract}
There are experimental indications of large 
flavor mixing between $\nu _{\mu }$ and $\nu _{\tau }$. 
In the unification models, 
in which the massless sector includes extra particles 
beyond the standard model, 
there possibly appear the mixings between quarks 
(leptons) and the extra particles. 
When large mixings occur, 
lepton flavor mixings can be quite different 
from quark flavor mixings. 
By taking the string inspired $SU(6) \times SU(2)_R$ 
model with global flavor symmetries, 
we obtain the neutrino flavor mixing 
$\sin \theta _{23} \simeq \lambda = \sin \theta _C$ 
around the unification scale. 
It can be expected that due to large Yukawa couplings 
of neutrinos, the renormalization effect increases 
$\sin 2 \theta _{23}$ naturally up to $\sim 1$ at the 
electroweak scale. 
Fermion mass spectra and the CKM matrix of quarks 
obtained in this paper are also phenomenologically viable. 
\end{abstract}

\newpage 
\section{Introduction}
The hierarchical patterns of quark-lepton masses and 
flavor mixings have been one of the outstanding 
problems in particle physics. 
From the viewpoint of quark-lepton unification 
it seems to be  plausible that the hierarchical 
structure of Yukawa couplings for leptons is similar 
to that for quarks and that lepton flavor mixings 
are also parallel to quark flavor mixings. 
Experimentally, however, the large neutrino flavor mixing 
$ \sin \theta _{23}$ has been suggested by the muon 
neutrino deficit in the atmospheric neutrino flux
\cite{Atmos}. 
This implies that lepton flavor mixings are remarkably 
different from quark flavor mixings in their 
hierarchical pattern. 
A natural question arises as to whether or not 
the distinct flavor mixings of quarks and leptons are 
in accord with the quark-lepton unification.

{}From the viewpoint of unification theory, 
it is reasonable that the hierarchical structure 
of Yukawa couplings is attributable to some kinds of 
the flavor symmetry at the unification scale $M_U$. 
If there exists the flavor symmetry such as ${\bf Z}_N$ or 
$U(1)$ in the theory, 
it is natural that Froggatt-Nielsen mechanism is 
at work for the interactions
\cite{Frog}. 
For instance, the effective Yukawa interactions for 
up-type quarks are of the form 
\begin{equation}
  M_{ij} \, Q_i U^c_j H_u 
\label{eqn:upq}
\end{equation}
with 
\begin{equation}
  M_{ij} = c_{ij} \left( \frac{\langle X \rangle}
                              {M_U}\right)^{m_{ij}} 
           = c_{ij} \, x^{m_{ij}}, 
\label{eqn:Mij}
\end{equation}
where subscripts $i$ and $j$ stand for the generation 
indices and all of the constants $c_{ij}$ are of order 
$O(1)$ with rank\,$c_{ij}=3$. 
The superfield $X$, which is singlet under the unification 
gauge group, is an appropriate composite superfield 
with the canonical normalization. 
The vaccum expectation value (VEV) of $X$ is 
supposed to be slightly smaller than $M_U$, 
where $M_U$ is nearly equal to the reduced Planck scale. 
For simplicity, we introduce global flavor $U(1)$ 
symmetry and ${\bf Z}_2$-symmetry ($R$-parity) 
at the unification scale. 
The charge of the superfield $X$ is assumed to be $(-1,+)$ 
under $U(1) \times {\bf Z}_2$. 
Instead of $U(1)$ we may take the ${\bf Z}_N$-symmetry. 
In that case the present analysis remains unchanged. 
Due to the $U(1)$-symmetry the exponents $m_{ij}$ 
in Eq.(\ref{eqn:Mij}) are determined according as 
the $U(1)$-charges of $Q_i$, $U^c_j$ and $H_u$. 
Here we denote the differences of $U(1)$-charges 
for $Q_2$-$Q_1$, $Q_3$-$Q_2$, $U^c_2$-$U^c_1$ and 
$U^c_3$-$U^c_2$ by $\alpha $, $\gamma $, $\beta $ 
and $\delta $, respectively. 
We have 
\begin{equation} 
    m_{ij} = m_{33} + \left(
      \begin{array}{ccc}
        \alpha + \beta + \gamma + \delta  &  
                \alpha + \gamma + \delta  &  
                      \alpha + \gamma   \\ 
        \beta + \gamma + \delta  &  
              \gamma + \delta  &  \gamma  \\ 
        \beta + \delta  &  \delta  &  0   \\
      \end{array} 
             \right)_{ij}, 
\label{eqn:mij} 
\end{equation} 
provided that $\alpha $, $\gamma $, $\beta $, $\delta $ 
and $m_{33}$ are non-negative. 
The mass matrix of the up-type quarks is described by 
the matrix $M$ multiplied by $v_u = \langle H_u \rangle $. 
By taking an ansatz that only top-quark has a 
trilinear coupling, i.e. 
\begin{equation}
  m_{33} = 0, 
\end{equation}
we obtain mass eigenvalues 
\begin{equation}
   O(v_u \,x^{\alpha + \beta + \gamma + \delta }), \qquad 
   O(v_u \,x^{\gamma + \delta }), \qquad  O(v_u), 
\label{eqn:muct}
\end{equation}
which correspond to $u$-, $c$- and $t$-quarks,
respectively. 
Thus, naively, the mass hierarchy of quarks and leptons, 
up to the renormalization effects, seems to be 
controlled only by $U(1)$-charges of the matter fields. 
However, in a wide class of unification models, 
the situation is not so simple. 
This is because the massless sector in the unification 
theory includes extra particles beyond the standard model 
and then there may occur extra-particle mixings such as 
between quarks(leptons) and colored Higgs 
fields(doublet Higgs fields). 
In order to study fermion masses and flavor mixings 
we have to take the effects of the extra-particle 
mixings into account. 
In addition, in the neutrino sector we should incorporate 
the extra-particle mixings with the see-saw mechanism
\cite{Seesaw}. 
In Ref.\cite{QCKM} we explained the observed hierarchical 
structure of the Cabbibo-Kobayashi-Maskawa(CKM) matrix 
for quarks in the string inspired $SU(6) \times SU(2)_R$ 
model.

In this paper we explore the CKM matrix for leptons. 
Several authors have pointed out that large neutrino 
flavor mixing can be obtained as a consequence of the 
cooperation between Dirac and Majorana mass matrices, 
provided that the Majorana mass matrix has a specific 
structure
\cite{Enhance}. 
In the quark-lepton unification, however, the Majorana 
mass matrix is closely linked to the other mass matrices 
and then it is difficult to expect such a cooperation 
between the Dirac and the Majorana mass matrices. 
In this paper, on the basis of $E_6$-type 
unification models we show that neutrino 
flavor mixing $\sin \theta _{23}$ is nearly equal to 
$\lambda = \sin \theta _C$ as an initial condition 
around the unification scale. 
Due to large Yukawa couplings of neutrinos 
the renormalization effect increases $\sin \theta _{23}$ 
naturally up to $\sim 1$ at the electroweak scale
\cite{Babu}\cite{Tanimoto}. 
Although in the present study we take a specific string 
inspired model, 
our results are applicable to a wide class of 
the unification models.

This paper is organized as follows. 
In section 2 we explain the interrelation between 
the hierarchical structure of mass matrices and 
the flavor symmetries based on the string inspired 
$SU(6) \times SU(2)_R$ model. 
Section 3 contains the mass spectra and the CKM 
matrix of quarks, which have been obtained in Ref.
\cite{QCKM}. 
After solving extra-particle mixings including 
the see-saw mechanism, 
we derive the CKM matrix of leptons in section 4. 
In section 5 we discuss the renormalization effect of 
the CKM matrices from the unification scale to 
the electroweak scale. 
Section 6 is devoted to summary.

\section{The hierarchical structure of mass matrices}
In this study we choose $SU(6) \times SU(2)_R$ 
as the gauge symmetry at the unification scale, 
which can be derived from the superstring theory 
via the flux breaking
\cite{Aligned}. 
Under $SU(6) \times SU(2)_R$ doublet Higgs and 
color triplet Higgs fields transform differently. 
This situation is favorable to solve the triplet-doublet 
splitting problem
\cite{Weak}. 
The chiral superfields, if we represent them in terms of 
$E_6$, consist of 
\begin{equation} 
       N_f \,{\bf 27} \ 
               + \ \delta \,({\bf 27} + {\bf 27^*}) 
\end{equation} 
except for $E_6$-singlets, 
where $N_f$ denotes the family number at low energies 
and $\delta $ is the set number of vector-like multiplets. 
Let us now consider the case $N_f = 3$ and $\delta =1$. 
Under $SU(6) \times SU(2)_R$ chiral superfields in 
${\bf 27}$ representation of $E_6$ are decomposed into 
\begin{eqnarray*} 
\Phi ({\bf 15, 1})  &:& \ \ Q, L, g, g^c, S, \\
\Psi ({\bf 6^*, 2}) &:& \ \ U^c, D^c, N^c, E^c, 
                                      H_u, H_d, 
\end{eqnarray*} 
where $g$, $g^c$ and $H_u$, $H_d$ represent colored 
Higgs and doublet Higgs fields, respectively. 
$N^c$ stand for the right-handed neutrino superfield 
and $S$ is an $SO(10)$-singlet. 
Although $L$ and $H_d$ ($D^c$ and $g^c$) 
have the same quantum numbers under the standard model 
gauge group $G_{SM} = SU(3)_c \times SU(2)_L \times U(1)_Y$, 
they belong to different irreducible 
representations of $SU(6) \times SU(2)_R$. 
Gauge invariant trilinear couplings are of the forms 
\begin{eqnarray} 
    (\Phi ({\bf 15, 1}))^3 & = & QQg + Qg^cL + g^cgS, \\
    \Phi ({\bf 15, 1})(\Psi ({\bf 6^*, 2}))^2 & 
            = & QH_dD^c + QH_uU^c + LH_dE^c  + LH_uN^c 
                                            \nonumber \\ 
             {}& & \qquad   + SH_uH_d + 
                     gN^cD^c + gE^cU^c + g^cU^cD^c.
\end{eqnarray} 

In the present model the massless matter fields are composed 
of chiral multiplets $\Phi _i$, $\Psi _i$ $(i=1,2,3)$ 
and one set of vector-like multiplets $\Phi _0$, 
$\Psi _0$ and ${\overline \Phi}$, ${\overline \Psi}$. 
Supposing that ordinary quarks and leptons are included 
in chiral multiplets $\Phi _i$, $\Psi _i$ $(i=1,2,3)$, 
$R$-parity of all $\Phi _i$, $\Psi _i$ $(i=1,2,3)$ 
are set to be odd. 
Since light Higgs scalars are even under $R$-parity, 
light Higgs doublets are bound to reside in 
$\Psi _0$ and/or ${\overline \Psi}$.
For this reason we assign even $R$-parity to 
vector-like multiplets. 
It is expected that the $R$-parity remains unbroken 
down to the electroweak scale
\cite{QCKM}. 
Hereafter we use the notations $\alpha$, $\beta$, 
$\gamma$ and $\delta$ defined by 
\begin{equation} 
    a_2 - a_1  \equiv \alpha , \quad 
    b_2 - b_1  \equiv \beta ,  \quad
    a_3 - a_2  \equiv \gamma , \quad 
    b_3 - b_2  \equiv \delta , 
\end{equation} 
where $a_i$ and $b_i$ $(i=1,2,3)$ 
represent the global $U(1)$-charges of 
matter fields $\Phi _i$ and $\Psi _i$, respectively. 
All of $\alpha $, $\beta $, $\gamma $ and $\delta $ 
are taken to be positive.

As shown in Ref.\cite{QCKM}, under an appropriate 
condition the gauge symmetry $SU(6) \times SU(2)_R$ 
is spontaneouly broken in two steps at the scales 
$\langle \Phi _0 \rangle = \langle {\overline \Phi} 
\rangle $ ($\langle S_0 \rangle = \langle {\overline 
S \rangle }$) and 
$\langle \Psi _0 \rangle = \langle {\overline \Psi} 
\rangle $ ($\langle N^c_0 \rangle = \langle {\overline 
N^c \rangle }$) as 
\begin{eqnarray} 
   SU(6) \times SU(2)_R 
   & \buildrel \langle S_0 \rangle \over \longrightarrow &
             SU(4)_{PS} \times SU(2)_L \times SU(2)_R \\
   & \buildrel \langle N^c_0 \rangle \over \longrightarrow &
             SU(3)_c \times SU(2)_L \times U(1)_Y, 
\end{eqnarray} 
where $SU(4)_{PS}$ stands for the Pati-Salam $SU(4)$
\cite{Pati}. 
Further, when $N^c_0 {\overline N^c}$ has the same 
$U(1)$-charge as $(S_0 {\overline S})^k$ 
for a positive integer $k$, 
we are led to the relation 
$(\langle S_0 \rangle /M_U)^k \simeq 
\langle N^c_0 \rangle /M_U$ \cite{Majorana}. 
We will assume that Froggatt-Nielsen mechanism 
is at work and that the superfields $X$ in 
Eq.(\ref{eqn:Mij}) is expressed in terms of 
$\Phi _0$, ${\overline \Phi}$ and/or moduli fields. 
In our scheme this assumption is natural. 
The $SU(6) \times SU(2)_R$ invariant and 
the global $U(1) \times {\bf Z}_2$ invariant 
interactions which contribute to the mass matrices 
of quarks and leptons are classified into six types 
of the superpotential term 
\begin{eqnarray}
   & & \Phi _i \Phi _j \Phi _0 X^{z_{ij}}, \, 
   \qquad \  \Phi _i \Psi _j \Psi _0 X^{m_{ij}}, 
   \qquad \, \Psi _i \Psi _j \Phi _0 X^{h_{ij}}, \\
   & & (\Phi _i {\overline \Phi})
            (\Phi _j {\overline \Phi}) X^{s_{ij}}, 
   \ \ (\Phi _i {\overline \Phi})
            (\Psi _j {\overline \Psi}) X^{t_{ij}},  
   \ \ (\Psi _i {\overline \Psi})
            (\Psi _j {\overline \Psi}) X^{n_{ij}} 
\end{eqnarray}
in $M_U$ units with $i,\ j = 1,2,3$. 
The exponents are expressed as 
\begin{eqnarray} 
  { }& &  z_{ij} - z_{33} = s_{ij} - s_{33} = \left(
      \begin{array}{ccc}
        2 \alpha + 2 \gamma  &  \alpha + 2 \gamma  &  
                      \alpha + \gamma   \\ 
        \alpha + 2 \gamma  &  2 \gamma  &  \gamma  \\ 
        \alpha + \gamma  &  \gamma  &  0   \\
      \end{array} 
             \right)_{ij},  \\ 
  { }& &  m_{ij} = t_{ij} -t_{33} = \left(
      \begin{array}{ccc}
        \alpha + \beta + \gamma + \delta  &  
                \alpha + \gamma + \delta  &  
                      \alpha + \gamma   \\ 
        \beta + \gamma + \delta  &  
              \gamma + \delta  &  \gamma  \\ 
        \beta + \delta  &  \delta  &  0   \\
      \end{array} 
             \right)_{ij}, \\ 
  { }& &  h_{ij} - h_{33} = n_{ij} - n_{33} = \left(
      \begin{array}{ccc}
        2 \beta + 2 \delta  &  \beta + 2 \delta  &  
                      \beta + \delta   \\ 
        \beta + 2 \delta  &  2 \delta  &  \delta  \\ 
        \beta + \delta  &  \delta  &  0   \\
      \end{array} 
             \right)_{ij}, 
\end{eqnarray} 
provided that $z_{33}, \ s_{33}, \ t_{33}, 
\ h_{33}, \ n_{33} > m_{33} = 0$. 
At energies below the scale 
$\langle N^c_0 \rangle (\equiv M_{PS} = M_U \,y)$ 
we have six types of bilinear terms 
\begin{eqnarray}
  & & M_U \,Z_{ij} \Phi _i \Phi _j, \ \  
      M_U \,G_{ij} \Phi _i \Psi _j, \ \  
      M_U \,H_{ij} \Psi _i \Psi _j, \\ 
  & & M_U \,S_{ij} \Phi _i \Phi _j, \ \  
      M_U \,T_{ij} \Phi _i \Psi _j, \ \ \, 
      M_U \,N_{ij} \Psi _i \Psi _j 
\end{eqnarray}
with 
\begin{eqnarray}
   Z & = & y_Z \Gamma _1 Z_0 \Gamma _1, \ \ \, 
    G  =   y \,\Gamma _1 G_0 \Gamma _2, \ \ \ \, 
    H  =   y_H \Gamma _2 H_0 \Gamma _2, \\
   S & = & y_S \Gamma _1 S_0 \Gamma _1, \ \ \  
    T  =   y_T \Gamma _1 T_0 \Gamma _2, \ \ \  
    N  =   y_N \Gamma _2 N_0 \Gamma _2, 
\end{eqnarray}
where the matrices $\Gamma _1$ and $\Gamma _2$ are 
defined by 
\begin{eqnarray}
  \Gamma _1 & = & 
      diag(\,x^{\alpha + \gamma }, \ x^{\gamma }, \ 1), \\
  \Gamma _2 & = & 
      diag(\,x^{\beta + \delta }, \ x^{\delta }, \ 1)
\end{eqnarray}
and all elements of the rank 3 matrices $Z_0$, $G_0$ 
and so on are of order $O(1)$. 
We suppose that there is no fine-tuning among elements 
of each matrix $Z_0$, etc. 
The normalization constants $y_Z$, $y_H$, $y_S$, $y_T$ 
and $y_N$ are expressed in terms of VEVs of 
$G_{SM}$-neutral fields divided by $M_U$. 
We have the relation $G = y M$. 
In what follows we direct our attention to the case 
\begin{eqnarray}
     M_U \,y, \ M_U \,y_Z, \ M_U \,y_H & = & 
                          10^{16} \sim 10^{17} \,{\rm GeV}, \\
     M_U \,y_S, \ M_U \,y_T, \ M_U \,y_N & = & 
                          10^{11} \sim 10^{12} \,{\rm GeV}. 
\end{eqnarray}

\section{The CKM matrix of quarks}
In this section we briefly review the results 
obtained in Ref.\cite{QCKM}. 
The mass matrix $M$ of the up-type quarks can be 
diagonalized by a bi-unitary transformation as 
\begin{equation} 
    {\cal V}_u^{-1} M \,{\cal U}_u. 
\end{equation} 
Since we have the form 
\begin{equation}
    M = \Gamma_1 G_0 \Gamma _2, 
\end{equation}
${\cal V}_u$ and ${\cal U}_u$ become 
\begin{eqnarray} 
    {\cal V}_u & = & 
      \left( 
      \begin{array}{ccc}
        1 - O(x^{2\alpha })  &  O(x^{\alpha})  
                          &  O(x^{\alpha + \gamma })  \\
        O(x^{\alpha })    &  1 - O(x^{2\alpha })  
                                   &  O(x^{\gamma })  \\
        O(x^{\alpha + \gamma })   &  O(x^{\gamma })  
                              &  1 - O(x^{2\gamma })  \\
      \end{array}
      \right), \\
    {\cal U}_u & = & 
      \left( 
      \begin{array}{ccc}
        1 - O(x^{2\beta })  &  O(x^{\beta})  
                          &  O(x^{\beta + \delta })  \\
        O(x^{\beta })    &  1 - O(x^{2\delta })  
                                   &  O(x^{\delta })  \\
        O(x^{\beta + \delta })   &  O(x^{\delta })  
                              &  1 - O(x^{2\delta })  \\
      \end{array}
      \right). 
\end{eqnarray} 
The mass eigenvalues are given in Eq.(\ref{eqn:muct}).

For down-type quarks there appear the mixings between 
$g^c$ and $D^c$ at energies below the scale 
$M_{PS} = \langle N_0^c \rangle$. 
An early attempt of explaining the CKM matrix via 
$D^c$-$g^c$ mixings has been made in Ref.\cite{SO10}, 
in which a SUSY $SO(10)$ model was considered. 
The mass matrix of down-type colored fields is 
written as 
\begin{equation} 
\begin{array}{r@{}l} 
   \vphantom{\bigg(}   &  \begin{array}{ccc} 
          \quad   g^c   &  \  D^c  &  
        \end{array}  \\ 
\widehat{M}_d = 
   \begin{array}{l} 
        g   \\  D  \\ 
   \end{array} 
     & 
\left( 
  \begin{array}{cc} 
       Z    &     G     \\
       0    &  \rho _d M 
  \end{array} 
\right) 
\end{array} 
\label{eqn:Mdh} 
\end{equation} 
in $M_U$ units, where we used the notation 
$\rho _d = \langle H_{d0} \rangle /M_U = v_d /M_U$. 
Note that $G = y M$. 
Since $\rho _d$ is a very small number($\sim 10^{-16}$), 
the left-handed light quarks consist almost only of 
$D$-components of the quark doublet $Q$. 
On the other hand, the mixing between $g^c$ and $D^c$ 
can be sizeable depending on the ratio $y_G/y$. 
The mixing matrix ${\cal V}_d$ of the $SU(2)_L$-doublet 
light quarks $D$ is determined such that 
${\cal V}_d^{-1}(A_d^{-1} + B_d^{-1})^{-1}{\cal V}_d$ 
becomes diagonal, 
where $A_d$ and $B_d$ stand for $Z Z^{\dag}$, 
$G G^{\dag}$, respectively. 
From the expression 
\begin{equation}
   (A_d^{-1} + B_d^{-1})^{-1} = \Gamma _1 
       \left( y_Z^{-2}(Z_0 \Gamma _1^2 Z_0^{\dag})^{-1} 
       + y^{-2}(M_0 \Gamma _2^2 M_0^{\dag})^{-1} 
        \right)^{-1} \Gamma _1, 
\end{equation}
we find that corresponding elements of 
the matrices ${\cal V}_d$ and ${\cal V}_u$ are of 
the same order of magnitude. 
Let us assume that the (1,1) elements of $A_d^{-1}$ 
and $B_d^{-1}$ are of the same order, i.e. 
\begin{equation}
   y_Z \, x^{\alpha + \gamma } \simeq y \, x^{\beta + \delta }. 
\end{equation}
In this case the coefficients of the leading term in 
off-diagonal elements of ${\cal V}_u$ and ${\cal V}_d$ 
become different because of large mixing between 
$D^c$ and $g^c$. 
Consequently, the CKM matrix for quarks is given by 
\begin{equation} 
   V_{CKM}^Q = {\cal V}_u^{-1} {\cal V}_d = \left(
      \begin{array}{ccc}
        1 - O(x^{2\alpha })  &  O(x^{\alpha })  &  
                              O(x^{\alpha + \gamma })   \\ 
        O(x^{\alpha })  &  1 - O(x^{2\alpha })  &  
                              O(x^{\gamma })            \\ 
        O(x^{\alpha + \gamma })  &   O(x^{\gamma }) & 
                              1 - O(x^{2\gamma })    \\
      \end{array} 
             \right).
\label{eqn:CKM} 
\end{equation} 
For the down-type light quarks their masses squared 
are described as the eigenvalues of the matrix
\begin{equation}
    \epsilon _d^2 \times (A_d^{-1} + B_d^{-1})^{-1}, 
\end{equation}
with $\epsilon _d = \rho _d /y$. 
Thus, we have the mass eigenvalues 
\begin{equation}
  O(v_d\,x^{\alpha + \beta + \gamma + \delta }), \qquad 
  O(v_d\,x^{\beta + \gamma + \delta }), \qquad 
  O(v_d\,x^{-\alpha + \beta + \delta }) 
\end{equation}
at the scale $M_{PS}$, 
which correspond to $d$-, $s$- and $b$-quarks, 
respectively. 
In this model $v_u/v_d = \tan \beta \sim 1$ 
is preferable. 
The masses of the down-type heavy quarks are 
$O(M_U \,y)$.

\section{The CKM matrix of leptons}
We now proceed to study the flavor mixing of leptons. 
In the lepton sector large $L$\,-$H_d$ mixing possibly 
occurs at energies below the scale $M_{PS}$. 
It is worth emphasizing that both $L$ and $H_d$ are 
$SU(2)_L$-doublets. 
This situation is in contrast to the $D^c$-$g^c$ mixing. 
For charged leptons we obtain a $6 \times 6$ mass matrix 
\begin{equation} 
\begin{array}{r@{}l} 
   \vphantom{\bigg(}   &  \begin{array}{ccc} 
          \   H_u^+   &   E^{c+}  &  
        \end{array}  \\ 
\widehat{M}_l = 
   \begin{array}{l} 
        H_d^-  \\  L^-  \\ 
   \end{array} 
     & 
\left( 
  \begin{array}{cc} 
       H    &    0       \\
       G    &  \rho _d M 
  \end{array} 
\right) 
\end{array} 
\label{eqn:Mlh} 
\end{equation} 
in $M_U$ units. 
The matrix $\widehat{M}_l$ can be diagonalized by 
a bi-unitary transformation as 
\begin{equation} 
    \widehat{\cal V}_l^{-1} \widehat{M}_l \, 
                        \widehat{\cal U}_l. 
\label{eqn:Ml} 
\end{equation} 
The unitary matrices $\widehat{\cal V}_l$ and 
$\widehat{\cal U}_l$ have the forms 
\begin{eqnarray} 
   \widehat{\cal V}_l & \simeq & \left( 
   \begin{array}{cc} 
      H {\cal W}_l \,(\Lambda _l^{(0)})^{-1}  
         &  -(H^{\dag})^{-1} {\cal V}_l 
                     \,\Lambda _l^{(2)}  \\
      G {\cal W}_l \,(\Lambda _l^{(0)})^{-1}   
         &  (G^{\dag})^{-1} {\cal V}_l 
                         \,\Lambda _l^{(2)} \\
   \end{array} 
                       \right), 
\label{eqn:hatvl}                              \\
   \widehat{\cal U}_l & \simeq & \left( 
   \begin{array}{cc} 
      {\cal W}_l   &  -\epsilon _d (A_l + B_l)^{-1} 
                                     B_l {\cal V}_l \\
      \epsilon _d B_l (A_l + B_l)^{-1} {\cal W}_l    &  
                                           {\cal V}_l 
   \end{array} 
                       \right) 
\label{eqn:hatul}
\end{eqnarray} 
in the $\epsilon _d$ expansion, 
where $A_l = H^{\dag} H$ and $B_l = G^{\dag}G$. 
${\cal W}_l$ and ${\cal V}_l$ are unitary matrices 
which diagonalize $(A_l + B_l)$ and 
$(A_l^{-1} + B_l^{-1})^{-1}$ as 
\begin{equation} 
  {\cal W}_l^{-1} (A_l + B_l) {\cal W}_l 
                     = (\Lambda _l^{(0)})^2, \qquad 
  {\cal V}_l^{-1} (A_l^{-1} + B_l^{-1})^{-1} 
                 {\cal V}_l = (\Lambda _l^{(2)})^2, 
\label{eqn:WlVl} 
\end{equation} 
where $\Lambda _l^{(0)}$ and $\Lambda _l^{(2)}$ are 
diagonal. 
Equation (\ref{eqn:hatul}) implies that 
the light $SU(2)_L$-singlet charged leptons are 
mainly $E^c$-components. 
From Eq.(\ref{eqn:hatvl}), the mass eigenstates of 
light $SU(2)_L$-doublet charged leptons are given by 
\begin{equation}
    \tilde{L}^{-} = \Lambda _l^{(2)} {\cal V}_l^{-1} 
                 ( - H^{-1} H_d^{-} + G^{-1} L^{-} ). 
\label{eqn:l-}
\end{equation}
Consequently, when the elements of $H^{-1}$ and $G^{-1}$ 
are comparable to each other, 
there occurs large mixing between $H_d^-$ and $L^-$. 
By inspecting the expression 
\begin{equation}
   (A_l^{-1} + B_l^{-1})^{-1} = \Gamma _2 
       \left( y_H^{-2}(H_0^{\dag} \Gamma _2^2 H_0)^{-1} 
       + y^{-2}(M_0^{\dag} \Gamma _1^2 M_0)^{-1} 
        \right)^{-1} \Gamma _2, 
\end{equation}
it is easy to see that ${\cal V}_l$ has the same 
hierarchical pattern as ${\cal U}_u$. 
Let us take an additional ansatz 
\begin{equation}
   y_H \, x^{\beta + \delta } \simeq 
                y \, x^{\alpha + \gamma + \xi } 
\end{equation}
with $\beta - \alpha \leq \xi > 0$. 
This implies that the large mixing occurs between 
$H_d^-$ and $L^-$. 
Thus we find that 
\begin{equation}
   \Lambda _l^{(2)} \simeq y \,x^{\gamma } \times 
            diag( \,x^{\alpha + \beta + \delta + \xi}, \ 
                  x^{\alpha + \delta},\  1) 
               = y \,x^{\gamma } \,\Gamma _2 \Gamma _3, 
\label{eqn:Ll2}
\end{equation}
where 
\begin{equation}
    \Gamma _3  = diag( \,x^{\alpha + \xi}, \ 
                              x^{\alpha },\  1). 
\end{equation}
Masses of light charged leptons turn out to be 
\begin{equation}
   O(v_d\, x^{\alpha + \beta + \gamma + \delta + \xi }), 
                                         \qquad 
   O(v_d\, x^{\alpha + \gamma + \delta }), \qquad 
   O(v_d\, x^{\gamma }) 
\label{eqn:lmass}
\end{equation}
at the scale $M_{PS}$, 
which correspond to $e$-, $\mu $- and $\tau $-leptons, 
respectively.

In the neutral lepton sector we have a $15 \times 15$ 
mass matrix 
\begin{equation} 
\begin{array}{r@{}l} 
   \vphantom{\bigg(}   &  \begin{array}{cccccc} 
          \quad \, H_u^0   &  \ \  H_d^0  &  \quad L^0  
                          &  \quad \  N^c   &  \quad  S  &
        \end{array}  \\ 
\widehat{M}_{NS} = 
   \begin{array}{l} 
        H_u^0  \\  H_d^0  \\  L^0  \\  N^c  \\  S  \\
   \end{array} 
     & 
\left( 
  \begin{array}{ccccc} 
       0     &     H     &     G^T     
                       &      0     &  \rho _d M^T  \\
       H     &     0     &      0      
                       &      0     &  \rho _u M^T  \\
       G     &     0     &      0      
                       &  \rho _u M &       0       \\
       0     &     0     & \rho _u M^T 
                       &      N     &      T^T      \\
   \rho _d M & \rho _u M &      0      
                       &      T     &       S       \\
  \end{array} 
\right) 
\end{array} 
\label{eqn:Mns} 
\end{equation} 
in $M_U$ units, where $\rho _u = v_u/M_S$. 
The $6 \times 6$ submatrix 
\begin{equation}
   \widehat{M}_M = \left(
   \begin{array}{cc}
        N    &   T^T  \\
        T    &    S     
   \end{array}
   \right) 
\end{equation}
represents the Majorana mass terms. 
In the study of lepton flavor mixing, 
instead of the matrix (\ref{eqn:Mns}) 
we may consider the $12 \times 12$ submatrix 
\begin{equation} 
\begin{array}{r@{}l} 
   \vphantom{\bigg(}   &  \begin{array}{ccccc} 
          \  H_u^0   &   H_d^0  &  \  L^0  
                          &  \quad \  N^c   &
        \end{array}  \\ 
\widehat{M}_N = 
   \begin{array}{l} 
        H_u^0  \\  H_d^0  \\  L^0  \\  N^c  \\
   \end{array} 
     & 
\left( 
  \begin{array}{cccc} 
       0   &   H   &     G^T     &      0       \\
       H   &   0   &      0      &      0       \\
       G   &   0   &      0      &  \rho _u M   \\
       0   &   0   & \rho _u M^T &      N       \\
  \end{array} 
\right). 
\end{array} 
\label{eqn:Mn} 
\end{equation} 
We will show shortly that the result remains unchanged. 
By recalling the above study in the charged lepton sector, 
it is easy to see that the unitary matrix 
$\widehat{\cal U}_N$ which diagonalizes $\widehat{M}_N$ 
is of the form 
\begin{equation}
   \widehat{\cal U}_N \simeq 
   \left( 
   \begin{array}{ccc}
      {\cal W}_l  &      0      &      0       \\
           0      &  \widehat{\cal V}_l  &  0  \\
           0      &      0      &  {\cal U}_N   
   \end{array}
   \right) 
   \times 
   \left( 
   \begin{array}{cccc}
      1/\sqrt 2 & -1/\sqrt 2 &     0     &   0   \\
      1/\sqrt 2 &  1/\sqrt 2 &     0     &   0   \\
         0      &      0     &  {\cal V} &   0   \\
         0      &      0     &     0     &   1   
   \end{array}
   \right). 
\end{equation}
Note that $\widehat{\cal V}_l$ is a $6 \times 6$ matrix 
Eq.(\ref{eqn:hatvl}) and that ${\cal W}_l$ is defined 
in Eq.(\ref{eqn:WlVl}). 
The unitary matrix ${\cal U}_N$ diagonalizes the Majorana 
mass matrix $N$ as 
\begin{equation}
    {\cal U}_N^{-1} N {\cal U}_N 
               \simeq  M_U y_N \,(\Gamma _2)^2 
                 =     M_U y_N \times 
             diag( \, x^{2\beta + 2\delta }, \ 
                       x^{2\delta }, \ 1). 
\end{equation}
This means that the Majorana masses have also 
hierarchical structure. 
By examining the unitary matrix $\widehat{\cal U}_N$, 
it turns out that the light neutrino mass eigenstates 
are 
\begin{equation}
     \tilde{L}^0 = 
    {\cal V}^{-1} \Lambda _l^{(2)} {\cal V}_l^{-1} 
           ( - H^{-1} H_d^0 + G^{-1} L^0 ). 
\label{eqn:l0}
\end{equation}
Comparing these eigenstates $\tilde{L}^0$ with those of 
the light charged leptons $\tilde{L}^-$ given 
by Eq.(\ref{eqn:l-}), 
we find that ${\cal V}$ is the CKM matrix for leptons. 
${\cal V}$ is determined such that 
\begin{equation}
   {\cal V}^{-1} (\Lambda _l^{(2)} {\cal V}_l^{-1} 
           N^{-1} {\cal V}_l \Lambda _l^{(2)}) {\cal V} 
\label{eqn:VLVN}
\end{equation}
becomes diagonal. 
Using Eq.(\ref{eqn:Ll2}), 
we obtain 
\begin{eqnarray}
    \Lambda _l^{(2)} {\cal V}_l^{-1} 
           N^{-1} {\cal V}_l \Lambda _l^{(2)} 
      & \simeq &  \frac {y^2}{y_N} \, x^{2\gamma} \, 
             \Gamma _3 ( \Gamma _2 {\cal V}_l^{-1} 
             \Gamma _2^{-1} N_0^{-1} \Gamma _2^{-1} 
             {\cal V}_l \Gamma _2) \Gamma _3  \nonumber \\
      & = & \frac {y^2}{y_N} \, x^{2\gamma} \times 
      \left( 
      \begin{array}{ccc}
          O(x^{2\alpha + 2\xi }) & O(x^{2\alpha + \xi }) 
                               & O(x^{\alpha + \xi }) \\
          O(x^{2\alpha + \xi })  & O(x^{2\alpha })       
                               & O(x^{\alpha })       \\
          O(x^{\alpha + \xi })   & O(x^{\alpha })   
                               & O(1)   
      \end{array}
      \right). 
\label{eqn:alxi}
\end{eqnarray}
Consequently, the lepton flavor mixing matrix is of 
the form 
\begin{equation}
     V_{CKM}^L = {\cal V} = 
     \left( 
         \begin{array}{ccc}
         1-O(x^{2\xi })    &    O(x^{\xi })    
                       &  O(x^{\alpha + \xi }) \\
           O(x^{\xi})      & 1-O(x^{2\alpha }) 
                       &     O(x^{\alpha })    \\
      O(x^{\alpha + \xi }) &   O(x^{\alpha })   
                       &  1-O(x^{2\alpha })  
         \end{array}
     \right) 
\label{eqn:LCKM}
\end{equation}
at the scale $M_{PS}$. 
It should be emphasized that we obtain 
an interesting relation between the lepton mixing angle 
and the quark mixing angle 
\begin{equation}
     \sin \theta _{23}^L \simeq \sin \theta_{12}^Q  
\label{eqn:theta23}
\end{equation}
at the scale $M_{PS}$. 
The light neutrino masses are given by the eigenvalues 
of the matrix Eq.(\ref{eqn:alxi}) multiplied 
by $(\rho _u/y)^2$. 
Thus we have 
\begin{equation}
     m_{\nu _i} \simeq \frac {v_u^2}{M_U y_N} x^{2\gamma }
                               \, (\Gamma _3)^2 
          = \frac {v_u^2}{M_U y_N} x^{2\gamma }
        \times (\, x^{2\alpha + 2\xi }, \ x^{2\alpha }, \ 1). 
\end{equation}

In the case of the matrix Eq.(\ref{eqn:Mns}), 
$N^{-1}$ in Eq.(\ref{eqn:VLVN}) should be replaced by 
\begin{equation}
  \left(
  \begin{array}{cc}
     1,   &  -y H^{-1} M^T 
  \end{array}
  \right) \widehat{M}_M^{-1} 
  \left(
  \begin{array}{c}
              1            \\
     -y M (H^{\dag})^{-1}  
  \end{array}
  \right). 
\end{equation}
Since this quantity has essentially the same form as 
$N^{-1} \propto \Gamma _2^{-1} N_0^{-1} \Gamma _2^{-1}$, 
we obtain the same mixing matrix Eq.(\ref{eqn:LCKM}).

\section{The CKM matrices at the electroweak scale}
In the above study we obtain the CKM matrices for quarks 
and leptons at the scale $M_{PS}$. 
Although we derive a remarkable relation, i.e. 
Eq.(\ref{eqn:theta23}) at the scale $M_{PS}$, 
we should correctly relate them to the values at low 
energies where they are measured. 
Since matter fields of the third generation have 
large Yukawa couplings, 
it is important to investigate the renormalization 
effects of the flavor mixings $\sin \theta _{23}^Q$ 
and $\sin \theta _{23}^L$ from the scale $M_{PS}$ 
to the electroweak scale. 
Here $M_{PS}$ is supposed to be $10^{16} \sim 10^{17}$GeV
\cite{QCKM}. 
In Refs.\cite{Babu}\cite{Tanimoto}, it has been shown in 
the minimal supersymmetric standard model that 
the neutrino flavor mixing $\sin \theta _{23}^L$ 
increases significantly with running down to the electroweak 
scale by the renormalization group equation (RGE). 
From the scale $M_{PS}$ to the Majorana mass scale $M_U \,y_N$ 
we manipulate the RGE for Yukawa couplings. 
Below the scale $M_U \,y_N$ we study the RGE 
of the neutrino mass operator $\kappa $ \cite{Babu}, 
which is given by 
\begin{equation}
     \kappa = y^{-2} \times \Lambda _l^{(2)} {\cal V}_l^{-1} 
                     N^{-1} {\cal V}_l \Lambda _l^{(2)}. 
\label{eqn:kappa}
\end{equation}
Since effective Yukawa couplings of the first and 
the second generations are very small, 
the renormalization effect of $\sin \theta _{12}$ is also 
rather small compared with that of $\sin \theta _{23}$. 
In the lepton sector the $SU(3)_c$ gauge interaction 
does not contribute to the one-loop RGE of 
Yukawa couplings. 
Therefore, the renormalization effect of 
$\sin \theta _{23}^L$ becomes more significant compared 
with that of $\sin \theta _{23}^Q$. 
From rough estimations, we can expect that 
$\sin \theta _{23}$ is enhanced by a factor 2 in 
running down to the electroweak scale from 
the scale $M_{PS}$. 
In this case we have 
\begin{equation}
      \sin \theta _{23}^L(M_Z) \sim 
      2\sin \theta _{23}^L(M_{PS}) \simeq 
      2\sin \theta _{12}^Q(M_{PS}) \simeq 2\lambda 
\end{equation}
with $\lambda = \sin \theta _C \simeq 0.22$. 
This means that $\sin 2\theta _{23}^L(M_Z) = 0.8 \sim 0.9$. 
This is consistent with the atmospheric neutrino data
\cite{Atmos}. 
The detailed study of the renormalization effect of 
$\sin \theta _{23}$ will be presented elsewhere.

Confronting the CKM matrix for quarks obtained here 
with the observed one, 
it is feasible for us to take a simple parametrization 
\begin{equation}
   x^{\alpha } = \lambda =0.22, \qquad 
   \alpha = \delta : \beta = \gamma : \xi 
      = 1:2.5:1.5. 
\end{equation}
Under this parametrization the hierarchical 
fermion masses at the scale $M_{PS}$ become 
\begin{equation}
   v_u \times (O(\lambda ^7), \ O(\lambda ^{3.5}), 
                                        \ O(1)) 
\end{equation}
for up-type quarks, 
\begin{equation}
   v_d \times (O(\lambda ^7), \ O(\lambda ^6), 
                           \ O(\lambda ^{2.5})) 
\end{equation}
for down-type quarks, 
\begin{equation}
   v_d \times (O(\lambda ^{8.5}), \ 
         O(\lambda ^{4.5}), \ O(\lambda ^{2.5})) 
\end{equation}
for charged leptons. 
These results are consistent with the observed mass 
spectra of quarks and leptons. 
Further, the neutrino masses are 
\begin{equation}
     \frac {v_u^2}{M_U y_N} \times 
     (\, O(\lambda ^{10}), \ O(\lambda ^7), \ 
     O(\lambda ^5)) 
\end{equation}
at the scale $M_{PS}$. 
Taking $M_U \,y_N = 10^{11.5}\,$GeV as a typical 
Majorana mass scale, 
apart from the renormalization effect, we obtain 
\begin{equation}
    m_{\nu _e} \simeq 2 \times 10^{-4}{\rm eV}, \quad 
    m_{\nu _{\mu }} \simeq 4 \times 10^{-3}{\rm eV}, \quad 
    m_{\nu _{\tau }} \simeq 7 \times 10^{-2}{\rm eV}.  
\end{equation}
Provided that the renormalization effects 
of $\sin \theta _{23}^Q$ and $\sin \theta _{23}^L$ are 
roughly about a factor $2 \sim \lambda ^{-0.5}$, 
the CKM matrices at the electroweak scale are 
expressed as 
\begin{equation} 
   V_{CKM}^Q \sim \left(
      \begin{array}{ccc}
        1 - O(\lambda ^2)  &  O(\lambda )  &  
                              O(\lambda ^3)   \\ 
        O(\lambda )  &  1 - O(\lambda ^2)  &  
                              O(\lambda ^2)   \\ 
        O(\lambda ^3)  &   O(\lambda ^2) & 
                              1 - O(\lambda ^4)  
      \end{array} 
             \right) 
\label{eqn:CKMN} 
\end{equation} 
for quarks and 
\begin{equation} 
   V_{CKM}^L \sim \left(
      \begin{array}{ccc}
        1 - O(\lambda ^3)  &  O(\lambda ^{1.5})  & 
                              O(\lambda ^2)   \\ 
        O(\lambda ^{1.5})  &  1 - O(\lambda )    & 
                              O(\lambda ^{0.5})   \\ 
        O(\lambda ^2)      &  O(\lambda ^{0.5})  & 
                              1 - O(\lambda )   
      \end{array} 
             \right) 
\end{equation} 
for leptons. 
The hierarchical structure of $V_{CKM}^Q$ is also 
consistent with the experimental data. 
Furthermore, it is worthy of note that neutrino 
masses and mixings obtained in this paper roughtly 
accommodate both the small angle MSW solution to 
the solar neutrino problem
\cite{MSW}\cite{SAsol} as well as 
the $\nu _{\mu }$-$\nu _{\tau }$ oscillation solution 
to the muon neutrino deficit in the atmospheric 
neutrino flux. 
However, the present model cannot explain the LSND results 
which, if confirmed, indicate 
$\Delta m^2_{LSND} \sim 1 {\rm eV}^2$ and 
$\sin ^2 2\theta \sim 10^{-3}$
\cite{LSND}.

\section{Summary}
Large neutrino flavor mixing $\sin \theta _{23}^L$ 
has been suggested by the muon neutrino deficit in 
the atmospheric neutrino flux. 
This means that lepton flavor mixings are remarkably 
different from quark flavor mixings in their hierarchical 
pattern. 
The difference of the structure of 
the CKM matrix between quarks and leptons 
is explanable in the $SU(6) \times SU(2)_R$ model and 
in the $E_6$-type unification models. 
In the models the massless sector includes extra particles 
beyond the minimal supersymmetric standard model. 
Concretely, we have triplet Higgs and doublet 
Higgs fields in each generation. 
As a consequence of such matter contents, there appear 
the mixings between the down-type quarks $D^c$ and triplet 
Higgses $g^c$ and between doublet leptons $L$ and 
doublet Higgses $H_d$. 
These mixings, if large, turn the mass pattern of 
down-type quarks and leptons out of that of up-type 
quarks. 
Further, it is expected that the hierarchical structure 
of Yukawa couplings comes from Froggatt-Nielsen mechanism 
under some kind of the flavor symmetry. 
Combining the unification models with this mechanism, 
we obtained the CKM matrices $V_{CKM}^Q$ and $V_{CKM}^L$ 
which are remarkably different from each other in the 
hierarchical pattern. 
In particular, the relation 
\begin{equation}
      \sin \theta _{23}^L \simeq \sin \theta _{12}^Q 
\end{equation}
holds at the scale $M_{PS}$. 
Due to large Yukawa couplings of the third generation 
the renormalization effect of flavor mixing 
$\sin \theta _{23}$ is sizeable. 
It can be expected that neutrino flavor 
mixing $\sin \theta _{23}^L$ increases by a factor 
about 2 with running down to the electroweak scale 
by the RGE. 
Thus we obtain 
\begin{equation}
    \sin \theta _{23}^L \sim 2\lambda  = 0.4 \sim 0.5, 
\end{equation}
which is consistent with the data. 
Fermion mass spectra and the CKM matrix of quarks 
obtained here are also phenomenologically viable. 
The hierarchical structure of fermion masses and 
the CKM matrices of quarks and leptons provides 
an important clue to the matter contents and 
the flavor symmetries at the unification scale.

\vspace{2cm}

\section*{Acknowledgements}
The authors thank Prof. S. Kitakado for careful 
reading of the manuscript. 
This work is supported in part by the Grant-in-aid for 
Scientific Research, Ministry of Education, 
Science and Culture, Japan (No. 08640366).



\end{document}